\pdfoutput=1
\documentclass{JINST}
\usepackage{textcomp} 
\DeclareGraphicsExtensions{.pdf} 

\title{Level-1 jet trigger hardware for the ALICE electromagnetic calorimeter at LHC}

\author{O. Bourrion$^a$\thanks{Corresponding
author.}, R. Guernane$^a$, B. Boyer$^a$, J.L. Bouly$^a$ and G. Marcotte$^a$\\
\llap{$^a$}Laboratoire de Physique Subatomique et de Cosmologie,\\ 
Universit\'e Joseph Fourier Grenoble 1,\\
  CNRS/IN2P3, Institut Polytechnique de Grenoble,\\
  53, rue des Martyrs, Grenoble, France\\
  E-mail: \email{olivier.bourrion@lpsc.in2p3.fr}}

\abstract{The ALICE experiment at the LHC is equipped with an electromagnetic calorimeter (EMCal) designed to enhance its capabilities for jet measurement. In addition, the EMCal enables triggering on high energy jets. Based on the previous development made for the Photon Spectrometer (PHOS) level-0 trigger, a specific electronic upgrade was designed in order to allow fast triggering on high energy jets (level-1). This development was made possible by using the latest generation of FPGAs which can deal with the instantaneous incoming data rate of 26 Gbit/s and process it in less than 4 \textmu s.}


\begin{document}

\section{Overview}
The ALICE detector at the LHC (A Large Ion Collider Experiment) will carry out comprehensive measurements of high energy nucleus-nucleus collisions, in order to study the phase transition between confined matter and the Quark-Gluon Plasma (QGP). For this purpose, ALICE has been upgraded with a large acceptance electromagnetic calorimeter providing the neutral portion of the jet energy measurement and an efficient and unbiased trigger for high energy jets.

The calorimeter consists of 12288 towers of layered Pb-scintillator which are arranged in 2x2 towers (one module). 
EMCAL is composed of 10 regular supermodules and is completed with 2 thirds of supermodule. One regular SM is made of 24 strips of 12 modules and the thirds will be equipped with 24 strips of 4 modules.
\subsection{Supermodule electronics}
\begin{figure}
  \begin{center}
  \includegraphics[width=0.8\textwidth]{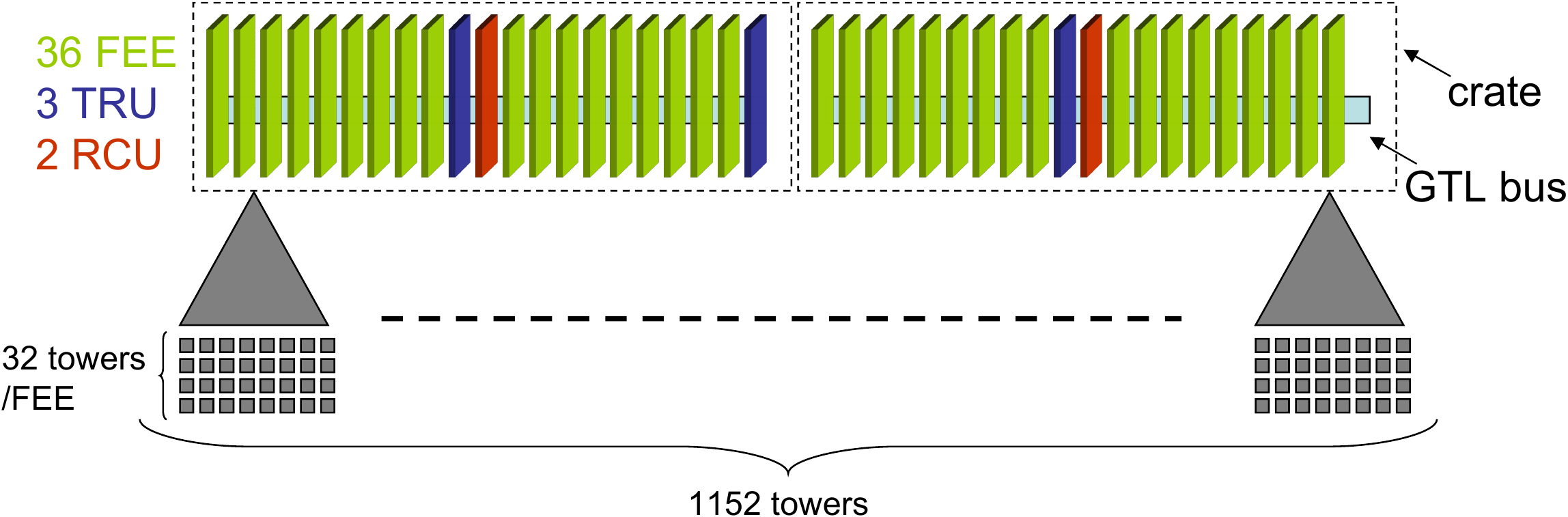} L
  \caption{One supermodule electronics}
  \label{SM_elec}
  \end{center}
\end{figure}

Each tower features a \textbf{C}harge \textbf{S}ensitive \textbf{P}reamplifier (CSP) coupled to an \textbf{A}valanche \textbf{P}hoto\textbf{D}iode (APD) that collects the light created during the interaction. The electronics necessary to read out one supermodule are distributed in 2 crates, see fig. \ref{SM_elec}. They are mostly occupied by front-end (FEE) cards \cite{FEEpaper}, whose primary purpose is to perform tower signal readout, thanks to the ALTRO chips \cite{ALTRObib}. Their secondary purpose is to build fastOR signals. Each of those is an analogue sum of the 4 CSP signals generated by a module that is fed through a fast shaper in order to minimize latency. The analogue fastOR signals are transfered to 3 \textbf{T}rigger \textbf{R}egion \textbf{U}nit (TRU) \cite{TRUpaper} which continuously digitize them at the machine bunch crossing rate (40 MHz) and compute the local L0 trigger. The \textbf{R}eadout \textbf{C}ontrol \textbf{U}nit (RCU) \cite{RCUpaper} is in charge of performing the data readout of both boards and transferring the data to the DAQ.

\subsection{TRU L0 algorithm}
After digitization, each fastOR is digitally integrated over a time sliding window of 4 samples. Then, the results of these operations are constantly fed to $2 \times 2$ spatial sum processors that compute the energy deposit in patches of $4\times4$ towers (or $2 \times 2$ fastOR), see fig. \ref{SMmap}. Each patch energy is constantly compared to a minimum bias threshold; whenever it is crossed and the maximum of the peak has been found, a local L0 trigger is fired. In preparation for the L1 algorithm, the time integrated sums are also stored in a circular buffer for later retrieval.

\section{Trigger requirements and hardware development motivations}
\begin{figure}
\begin{center} 
\includegraphics[width=0.3\textwidth,angle=-90]{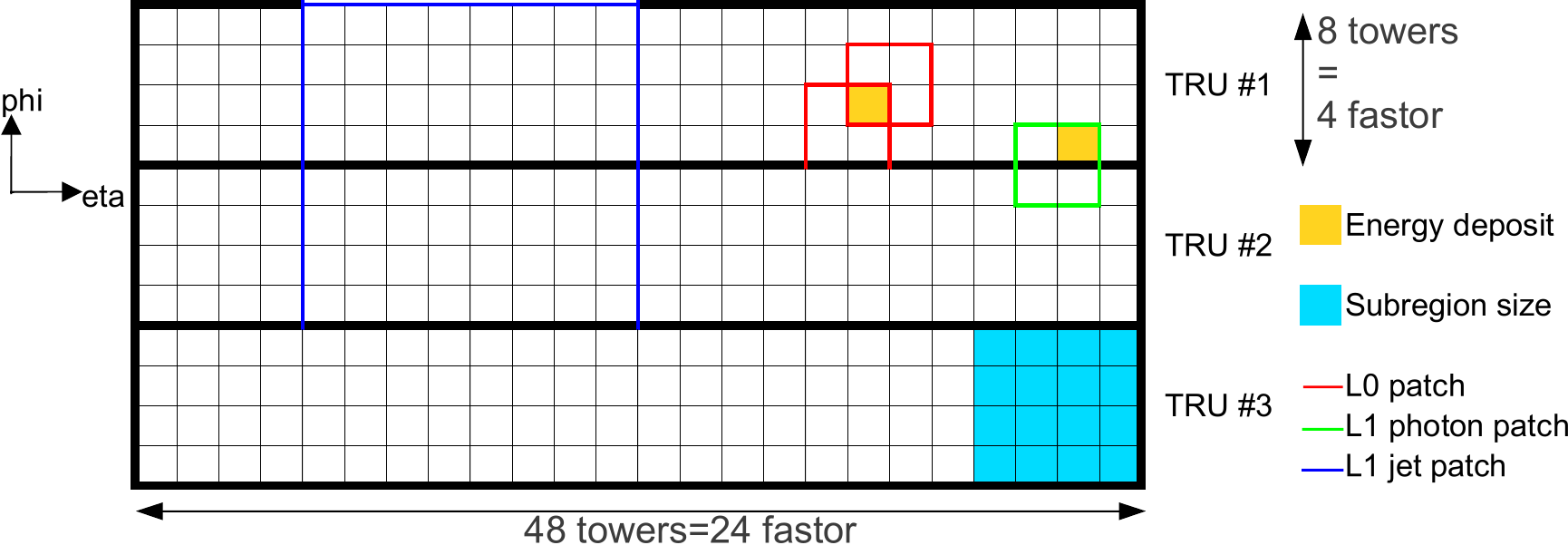}
\label{SMmap}
\caption{Flat view of one supermodule, constituted of three neighboring TRU regions, showing the size and arrangement of possible trigger patches. Each square depicts a module (fastOR) signal.}
\end{center}
\end{figure}

In order to meet the objective of recording, the required rejection is in the order of 10-20 for Pb+Pb and $\sim$3000 for p+p (smaller event size but higher collision rate). In the Pb+Pb case most of the rejection will be done by the \textbf{H}igh \textbf{L}evel \textbf{T}rigger (processor farm) as opposed to the p+p case, where it is ineffective and most of it must be done at L0/L1 level by hardware. A new hardware development was mandatory, first in order to build the global L0 trigger which is an OR of the 32 L0 locally calculated by the TRUs. Secondly, for computing two kinds of L1 triggers: the L1-gamma trigger which uses the same patch size as L0, but without the inefficiencies  displayed by the L0 (i.e. $2\times2$ patches across several TRU regions can be computed), and the L1-jet trigger, which is built by summing energy over a sliding window of $2\times2$ subregions (1 subregion = 4x4 fastOR = 8x8 towers) (see fig. \ref{SMmap}). For that, data aggregation is necessary.
Also, it is  desirable to maintain the L1 triggers selectivity against collision centrality in order to discriminate interesting signal from the background noise. The thresholds must therefore be corrected by the multiplicity information which is made available by the V0 detector \cite{V0paper}.

\section{Solution implemented: the Summary Trigger Unit (STU)}
\begin{figure}
\begin{center} 
\includegraphics[scale=0.5,angle=-90]{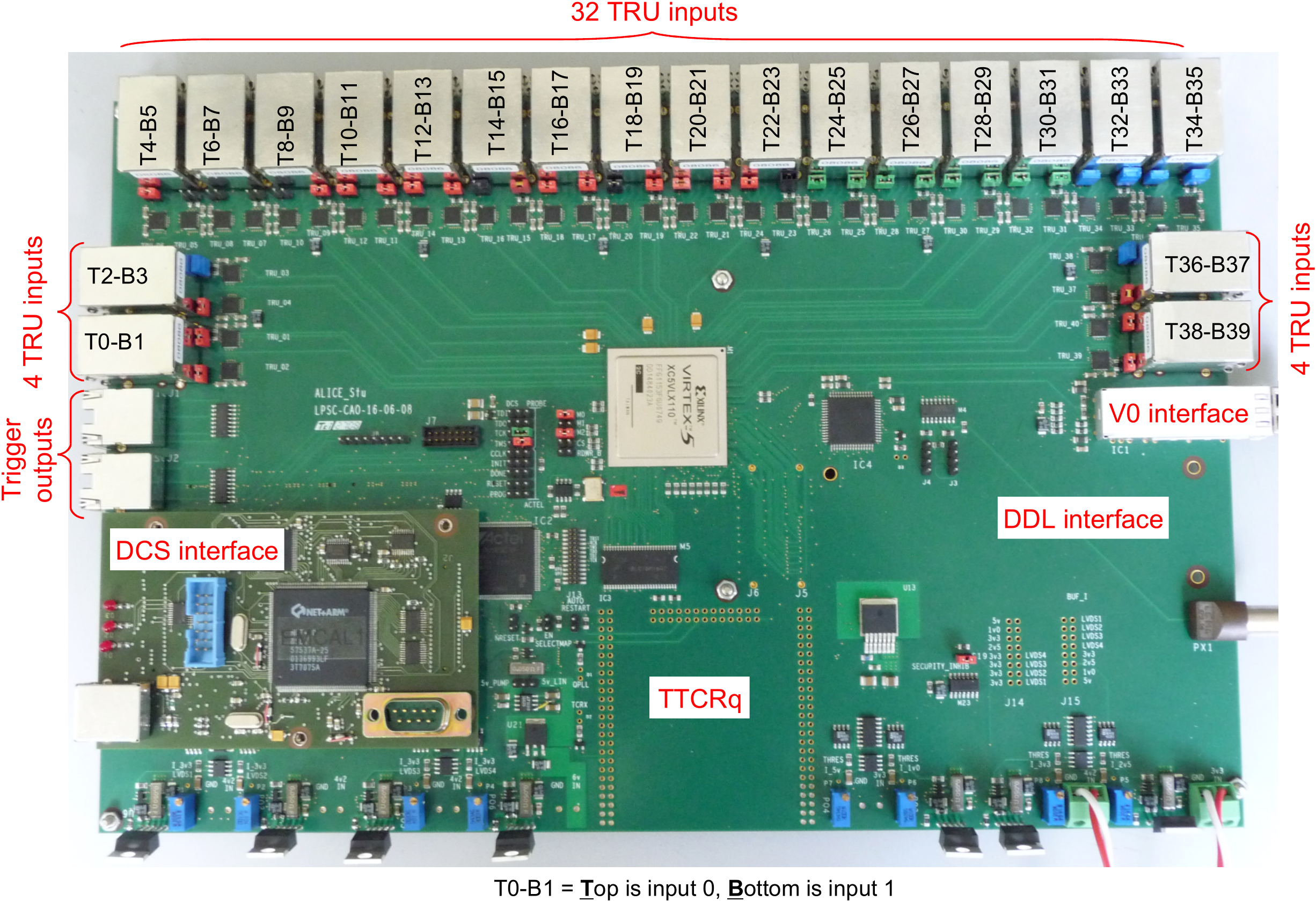}
\label{STU_overview}
\caption{Picture of the Summary Trigger Unit equipped with the xilinx XC5VLX110FF1153.}
\end{center}
\end{figure}

The STU hardware (fig. \ref{STU_overview}) is FPGA-based and is designed to allow data concentration, thanks to the custom serial link connections with all TRUs. STU also features a 
\textbf{T}rigger, \textbf{T}iming and \textbf{C}ontrol (TTC) interface \cite{TTCpaper} for receiving the machine reference clock and the trigger messages. The interface with V0 is there to allow event by event threshold computation according to a  second order fit of EMCAL energy as a function of V0 information (which is itself dependent on PMT HV): $A . V0^2 + B . V0+C$.
When a confirmed L2 trigger is received, STU data (TRU time sums, triggering patch position, thresholds used) are read out via a \textbf{D}etector \textbf{D}ata \textbf{L}ink (DDL) \cite{DDLpaper} on a per event basis.
For all of those, multievent buffering is implemented.
The \textbf{D}etector \textbf{C}ontrol \textbf{S}ystem (DCS) interface is based on a transformer-less Ethernet interface board allowing remote FPGA configuration and experiment configuration (thresholds, delays, ...).

\subsection{Custom serial protocol implementation}
The main motivation for the development of this custom serial link was the desire to reuse the TRU design made for the \textbf{PHO}ton \textbf{S}pectrometer (PHOS) which was equipped with a spare RJ45 connector directly linked to its FPGA.
Also, the trigger timing constraints drove the design in the same direction, because the transmission latency  minimization and some functional requirements, such as having the STU used as a low jitter reference clock distributor for TRUs and the fact that the local L0s had to be forwarded to STU for feeding its global OR required a proprietary solution. Thus, the choice was made to use a 4-pair LVDS link transported over CAT7 Ethernet cables because they have the appropriate impedance and feature low signal attenuation and low skew between pairs. Pair usage is as follows: one pair is dedicated for the LHC reference clock transfer to the TRU, another is used by the TRUs to forward their local L0 candidates and the 2 remaining are used for synchronous serial data transfer without any encoding. Each data pair running at 400 Mb/s and the clock used for transfer is the LHC clock multiplied by 10.
With this very light protocol, the latency is only  the sum of the cable delay and bit transmission time. Each TRU sends simultaneously its 96 values of 12 bit coded time integrated fastOR data to the STU at 800 Mb/s; the latency is thus $1.44\ \mu s$.

The link synchronization is done before each start of run by Finite State Machine implemented in the FPGA. It is done in 2 steps, first the data phase alignment takes place, it relies on the individual data path fine granularity delaying feature available in the Virtex 5 FPGA (up to 64 steps of 78 ps). A scanning of all tap values is made in order to obtain the zone where  data reception is stable and then the central value is applied. Secondly, character framing 
is performed for associating individually each incoming bit to the good deserialized word.

\subsection{Trigger algorithms description}
\begin{figure}
\begin{center}
\includegraphics[scale=0.6,angle=-90]{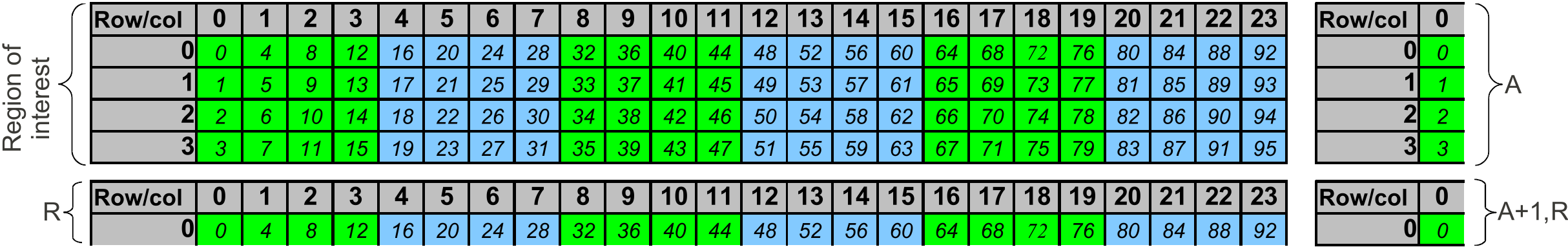}\\
\vspace{5pt}
\includegraphics[scale=0.6,angle=-90]{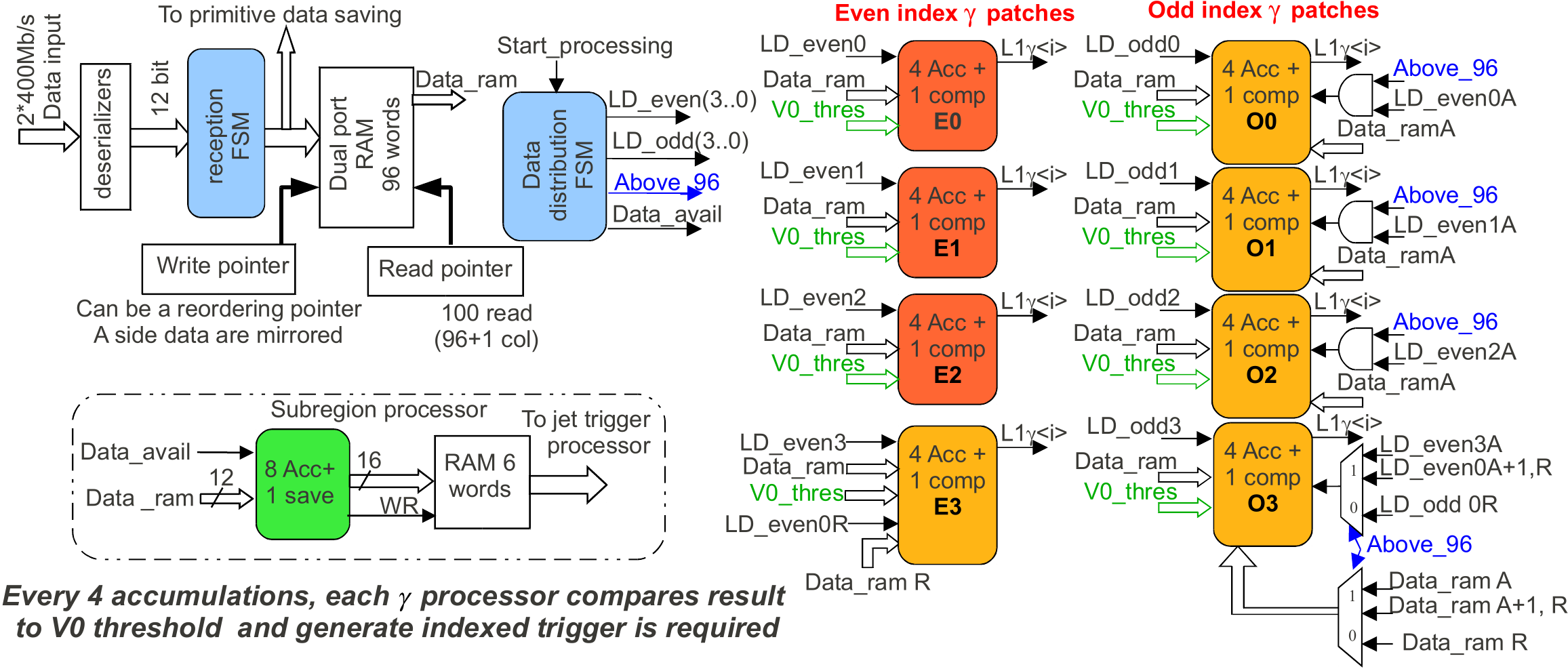}
\caption{On top, one TRU region with fastOR numbering, surrounded by its neighboring region. Below, one photon patches processor, containing deserializer, mirroring system, data distribution FSM, 2 columns of 4 photon processors and the subregion integrator.}
\label{photon_proc}
\end{center}
\end{figure}

The L1 trigger processing starts at the confirmed L0 reception on the RCU side. The trigger information is then passed to the TRUs which select the appropriate time integrated value in their circular buffers and transmit them to the STU. Meanwhile, the V0 detector transfers the A and C plate charge information to the STU via the direct optical link. The thresholds for photon and jet patches are immediately processed and made available before the actual patch processing starts.\\
 In the STU, one photon processor (fig. \ref{photon_proc}) is implemented by TRU reception link. For the A side, the data are mirrored as they are written in the reception buffer, i.e. fastOR \#95 is written in first memory position and fastOR \#0 in the last position. This compensates the actual physical supermodule mirroring due to their opposite sides of insertion.\\
Only 2 columns of 4 processors are needed per region, because there are 4 possible overlapping patches in $\phi$ and 2 in $\eta$. Each processor is an object doing 4 accumulations and one threshold comparison right after the fourth loading.
The processing of all regions is done synchronously and in parallel and data are dispatched to the photon processors according to their geographical position in the map. Also, some processors (orange in the fig. \ref{photon_proc}) are fed with the neighboring region data. For instance, patch processor E0 process [0, 1, 4, 5], [8, 9, 12, 13], ..., [88, 89, 92, 93] and O0 does [4, 5, 8, 9], ... ,[92, 93, 0(A), 1(A)].\\
In parallel to this processing, and for preparing the jet patch processing, data are also fed to a $4 \times 4$ integrator which builds the 6 subregions per region. This operation is basically only carrying out 6 times 16 successive accumulations followed by a memory write.
Once all subregion informations are built, EMCAL can be represented as a rectangle of 12 rows $\times$ 16 columns of subregions. At this stage only one jet processor is necessary. It is very similar to one of those used for photon processing in a region, due to the new geometry, 2 columns of 11 processors of $2 \times 2$ objects are needed (here the object being subregion instead of fastOR). 165 different patches are computed.

\section{Conclusion}
The STU has been installed for a year, all the system interfaces have been validated and the custom serial protocol has been demonstrated to operate in real condition with intensive readout. The L0/L1 triggers are commissioned and data validation with the runs done in p+p at fixed threshold is in progress.
The good trade-off between parallel and serial computation has allowed to use only $\sim$ 50\% of the FPGA internal logic and left a timing margin of $\sim$700 ns for the processing of 2961 photon $2\times2$ and 165 jet $64\times64$ patches processing. These provisions allow future algorithm upgrades or modifications.

\end{document}